\documentclass[letterpaper, 10 pt, journal, twoside]{ieeetran}
\IEEEoverridecommandlockouts                              
                                                          
\usepackage{epsfig}
\usepackage{url}
\usepackage{amsmath}
\usepackage{amsfonts}

\usepackage{amsthm}
\usepackage{amssymb}
\usepackage{graphicx}
\usepackage{color}
\usepackage{balance} 
\usepackage{algorithm2e}
\usepackage{algorithmic}
\usepackage{bm}

\newtheorem{corollary}{Corollary}

\newcommand{\bd}{\mathbf}
\DeclareMathOperator{\diag}{diag}

\usepackage{enumitem}
\usepackage{graphicx} 
\graphicspath{{figure/}}
\usepackage{float}
\usepackage[caption=false,font=footnotesize]{subfig}
\usepackage{soul}
\usepackage{color}

\usepackage{algorithm,algorithmic}
\usepackage{cite}

\allowdisplaybreaks
\usepackage{mathtools}

\newtheoremstyle{bfnote}%
{}{}%
{\itshape}{}%
{\bfseries}{.}%
{ }%
{\thmname{#1}\thmnumber{ #2}\thmnote{ (#3)}}
\theoremstyle{bfnote}
\newtheorem{thm}{Theorem}

\newtheorem{lem}{Lemma}
\newtheorem{rmk}{Remark}
\newtheorem*{base*}{Base Controller}

\usepackage{tikz}

\newcommand{\real}[0]{\mathbb R}

\DeclareSymbolFont{bbold}{U}{bbold}{m}{n}
\DeclareSymbolFontAlphabet{\mathbbold}{bbold}

\DeclarePairedDelimiterX\Set[2]{\lbrace}{\rbrace}%
{ #1 \,\delimsize| \,\mathopen{} #2 }

\usepackage{bookmark}

\title{\LARGE \bf Leveraging Predictions in Power System Voltage Control: An Adaptive Approach}

\author{Wenqi Cui, Yiheng Xie, Steven Low, Adam Wierman, Baosen Zhang
\thanks{Wenqi Cui, Yiheng Xie, Steven Low, Adam Wierman, is with the Department of Computing + Mathematical Sciences, California Institute of Technology, CA 91125, USA.  } %
\thanks{ Baosen Zhang  are with the Department of Electrical and Computer Engineering, University of Washington Seattle, WA 98195, USA zhangbao@uw.edu} 
\thanks{The authors are partially supported by the Resnik Sustainability Institute, the PIMCO Foundation, the NSF grants ECCS-1930605, 2200692, and 2153937.}}

\begin{document}
\maketitle
\thispagestyle{empty}
\pagestyle{empty}

\begin{abstract}
 High variability of solar PV and sudden changes in load (e.g., electric vehicles and storage) can lead to large voltage fluctuations in the distribution system.  In recent years, a number of controllers have been designed to optimize voltage control. These controllers, however, almost always assume that the net load in the system remains constant over a sufficiently long time, such that the control actions converge before the load changes again. Given the intermittent and uncertain nature of renewable resources, it is becoming important to explicitly consider net load that is time-varying.

This paper proposes an adaptive approach to voltage control in power systems with significant time-varying net load. We leverage advances in short-term load forecasting, where the net load in the system can be partially predicted using local measurements. We integrate these predictions into the design of adaptive controllers, and prove that the overall control architecture achieves input-to-state stability in a decentralized manner. We optimize the control policy through reinforcement learning. Case studies are conducted using time-varying load data from a real-world distribution system.

\end{abstract}



\section{Introduction}

The increasing adoption of distributed energy resources (DERs) brings high variability and intermittency in power injections across distribution systems. These fluctuations can lead to rapid and substantial voltage deviations that occur on timescales much shorter than those handled by conventional mechanical control devices such as tap-changing transformers~\cite{yuan2023learning, turitsyn2011options,bolognani2013distributed}. To address this issue, the fast-acting capabilities of inverter-based resources (IBRs) offer a promising alternative by controlling their reactive power injections in response to voltage deviations~\cite{yeh2012adaptive,zhang2014optimal,li2014real}. A plethora of control laws have been proposed for voltage regulation, and the most representative designs include linear droop control\cite{zhu2015fast, vandoorn2010active}, incremental voltage control law~\cite{farivar2015local, li2014real, qu2019optimal}, and learning-based monotone controllers~\cite{yuan2023learning, cui2022decentralized, feng2023stability, feng2023bridging}.

A common assumption underlying all of the aforementioned works is the existence of a separation in timescale, where the system has sufficient time to reach a steady state after a disturbance such as a sudden change of load. However, due to the intermittent and uncertain nature of renewable energy sources and larger compute loads, this assumption may no longer hold: net load variations can persist such that the system never reaches the steady state. Therefore, it becomes essential to explicitly model the time-varying characteristics of net load and to design voltage control laws that can effectively respond to these time-varying patterns.


One way to handle time variation is to use robust controllers, since they can potentially achieve performance guarantees under uncertainties.  To achieve robustness towards the uncertainties in PV generation,
\cite{jabr2019robust} proposes to modulate the
smart inverter reactive power as a function of its real power.  To address uncertainties over time horizons, a robust constrained model predictive control is constructed for centralized voltage control with respect to uncertain power outputs from DERs\cite{maharjan2020robust}. 
To mitigate the impact of uncertain grid topologies, \cite{yeh2022robust} and \cite{christakou2017voltage} propose robust voltage control mechanisms with an unknown grid topology. However, these robust control strategies often rely on real-time wide-area communication infrastructures, which may not be feasible in distribution networks. 
 To address communication limitations, \cite{liu2017distributed} proposes a distributed voltage control scheme using the alternating direction method of multipliers (ADMM) algorithm. Building upon the design in \cite{jabr2019robust}, \cite{shi2023data} proposes a data-driven and distributed scheme for robust optimization of voltage control gains, where system information and wide-area communications is not required. However, these approaches still require some form of information exchange among neighboring nodes, which can be difficult to implement in practice.  In addition, the robust formulation tends to produce conservative solutions that might be far from effective in typical operating scenarios.

Instead of modeling the time-varying load as uncertainties in a bounded region or following certain distributions, recent advances in short-term predictions show that future load or renewables can be largely predicted from historical data~\cite{xia2021stacked, wang2017data}. This motivates us to leverage local predictions of the net load in voltage control, so that the controller can adapt to time-varying changes of the net load.  However, prediction errors are inevitable, and local forecasts do not fully capture the influence of other buses in the network. This gives rise to two central challenges we aim to address: \textit{how to systematically embed predictions in control, and what performance guarantees can be established.}

\begin{figure*}[h]
    \centering
    \includegraphics[width=0.85\textwidth]{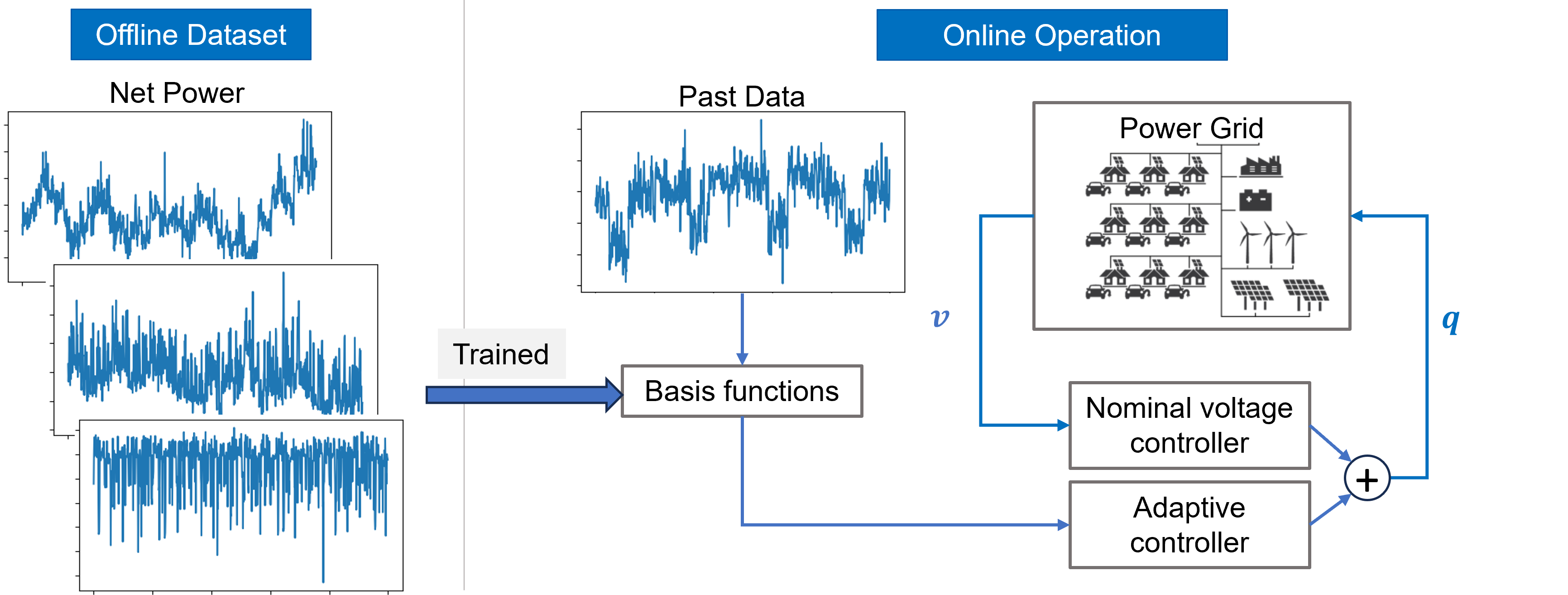}
    \caption{Structure of the adaptive approach for voltage control with time-varying net load. We consider predictions as a set of basis functions, and the predictable time-varying load is expressed through a linear combination of the basis functions. We integrate these predictions into the design of adaptive controllers, which provides a flexible way for incorporating diverse prediction models with minimal modifications to existing
voltage control laws. The overall control architecture achieves input-to-state stability in a decentralized manner.  }
    \label{fig:structure}
\end{figure*}

To address these challenges, this paper proposes to embed predictions in the voltage controller via an adaptive control framework. We consider predictions as a set of basis functions, and the predictable time-varying load is expressed through a linear combination of the basis functions. By designing an adaptive law on the combination coefficients, the predictable part is able to follow the trend of the real-time time-varying net load. This provides a flexible way for incorporating diverse prediction models with minimal modifications to existing voltage control laws. In addition, we show that the closed-loop control guarantees voltage stability despite the existence of prediction errors. Case studies are conducted using both illustrative net load and real-world load data from an operational distribution grid~\cite{xie2025digital}, demonstrating the effectiveness of the proposed approach in reducing voltage fluctuations and violations. 

The structure of the proposed approach is illustrated in Fig.~\ref{fig:structure}. The contributions of this paper is as follows:
\begin{enumerate}
    \item We propose an adaptive control framework to embed predictions into the voltage controller, which reduces voltage fluctuations in the presence of time-varying load and power injections.
    \item We prove that the proposed controller design guarantees the input-to-state stability of the voltage dynamics, where the impact of prediction error on voltage deviations is analytically quantified.
    \item The proposed method provides a framework for leveraging abundant offline data for real-time voltage control. Through the adaptive design, the proposed controller could bridge the gap between offline training of predictions and online operations.
\end{enumerate}

The remainder of this paper is organized as follows. Section II introduces the problem formulation for voltage control in distribution systems. Section III characterizes the modeling of time-varying net power injections and derives equivalent voltage dynamics for the design of decentralized controllers. Section IV proposes an adaptive approach to embed predictions with minimal modifications to existing voltage control laws. The conditions for achieving exponential input-to-state stability are also derived. Section V shows the simulation results, and Section VI concludes the paper.

\section{Model and Problem Formulation} \label{sec:model}
\subsection{Notation }
Throughout this manuscript, vectors are denoted in lower-case bold and matrices are denoted in upper-case bold, and scalars are unbolded unless otherwise specified. Vectors of all ones and zeros are denoted as  $\mathbbold{1}_n, \mathbbold{0}_n \in \real^n$, respectively. The identity matrix is denoted as $\mathbf{I}_n\in \real^n$.
Subscript $i$ indicates variables for the node $i$. For vectors $a_1, \cdots, a_n$, each associated with one of the $n$ nodes, $\bm{a}:=(a_1,\cdots, a_n)$ stacks them into a column vector, and $\hat{\bm{a}}:=\text{diag}(a_1,\cdots, a_n)$ arranges the variables into a diagonal matrix. Given a matrix $\mathbf{A}\in \real^{n\times n}$, $\diag(\mathbf{A})\in \real^{n\times n}$ is a diagonal matrix that extracts the diagonal part of $\mathbf{A}$, while $\overline{\diag}(\mathbf{A})\in \real^{n\times n}$ extracts the off-diagonal part of the matrix. The element of matrix $\mathbf{A}$ located in the $i$-th row and $j$-th column is written as $\mathbf{A}_{i,j}$.
The smallest and largest eigenvalues of the matrix $\mathbf{A}$ are $\lambda_{min}(\mathbf{A})$ and $\lambda_{max}(\mathbf{A})$, respectively. 
If a variable is a state in a dynamical system, the superscript $^*$ indicates its equilibrium. 


\subsection{Model }
A standard requirement for a distribution network is that the voltages should not be too far from their rated values (e.g., less than 5\%)~\cite{national1996american}. Following standard practice, we normalize the units so that the reference value for voltages at all buses is 1 per unit (p.u.)~\cite{zhang2014optimal,low2014convex}. 

For a power network with $n$ buses, let $\bm{v}(t):=\left(v_1(t),\cdots,v_n(t)\right)$ be the voltage vector where $v_i(t)$ is the voltage at bus $i$ at time $t$. Let $\bm{p}(t)$ be the vector of active power injections and $\bm{q}(t)$ be the vector of bus reactive power injections, respectively. We adopt the LinDistFlow model~\cite{baran2002optimal,zhou2020note,huang2021generalized}, which linearly relates the bus voltages to the power injections as 
\begin{equation}\label{eq:Dyn_voltage}
\bm{v}(t)=\mathbf{R} \bm{p}(t)+\mathbf{X} \bm{q}(t)+\mathbbold{1}_n,
\end{equation}
where $\mathbf{R}$ and $\mathbf{X}$ are positive definite matrices describing the network~\cite{zhu2015fast,farivar2013equilibrium}. 
In this paper, we assume that active power depends on the external environment, e.g., the power generation from distributed energy resources and time-varying power consumption. In contrast, reactive power is controllable,
for example, through inverter-based resources (IBRs)~\cite{zhang2014optimal}. 

As noted in the introduction, we explicitly consider the time dependence of $\bm{p}$ on $t$. That is, we do not assume that voltage regulation is fast enough so that voltages would reach their setpoint much faster than the variations in the active power injections. Nevertheless, the controllers should maintain the voltage close to their reference values despite these variations. 

Because of the lack of real-time communication in most distribution systems, we consider static local feedback controllers in this paper: bus $i$ determines the control action $u_i(t)$ based on its local measurements, e.g., the local voltage deviation $v_i(t)$. 
These controllers have been widely studied in the literature and a common structure is to update reactive power injections successively based on the local voltage measurements.\footnote{Note that a simple proportional control law, $q_i(t)=k_iv_i(t)$, cannot guarantee the convergence of voltage to its reference value at the steady state~\cite{zhang2014optimal}.} 
Denote $u_{i}(t)$ as the successive update of the reactive power for each bus, 
we update $\bm{q}$ and $\bm{v}$ iteratively as
\begin{subequations}\label{eq:Dynamic}
\begin{align}
\bm{q}(t\!+\!1)&=\bm{q}(t)-\bm{u}(t) , \label{subeq:Dynamic_q}\\
 \bm{v}(t\!+\!1)&=\mathbf{R} \bm{p}(t)+\mathbf{X} \left(\bm{q}(t)-\bm{u}(t)\right) + \mathbbold{1}_n, \label{subeq:Dynamic_v}
\end{align}
\end{subequations}
and the question becomes how to find good control law for $\bm{u}$. In the following, we write $\bm{u}_{\bm{\theta}}(t)$ as the control law parametrized by $\bm{\theta}$. The specific design for $\bm{u}_{\bm{\theta}}(t)$ will be explained in Section~\ref{sec:module}.

\subsection{Optimal voltage control}
Our objective is to optimize $\bm{u}_{\bm{\theta}}(t)$ to minimize cost in $\bm{v}$ and $\bm{q}$ over a time-horizon of length $T$. Let $C_q(\bm{q}(t))$ and $C_v(\bm{v}(t)-\mathbbold{1}_n) $ be the cost for reactive power and the voltage deviation, respectively. The optimization problem is 
\begin{subequations}\label{eq:Optimization}
\begin{align}
\min_{\bm{\theta}} & \quad \sum_{t=1}^{T} C_q(\bm{q}(t))+C_v(\bm{v}(t)-\mathbbold{1}_n) \label{subeq:Optimization_obj}\\
\mbox{s.t. } & \bm{q}(t\!+\!1)=\bm{q}(t)-\bm{u}_{\bm{\theta}}(t) \label{subeq:Optimization_q}\\
 &\bm{v}(t\!+\!1)=\mathbf{R} \bm{p}(t+1)+\mathbf{X} \left(\bm{q}(t)-\bm{u}_{\bm{\theta}}(t)\right) +\mathbbold{1}_n\label{subeq:Optimization_V}\\
& \bm{u}_{\bm{\theta}}(t)\text{ is stabilizing}\label{subeq:Optimization_stability}
\end{align}
\end{subequations}
where constraints~\eqref{subeq:Optimization_q}-\eqref{subeq:Optimization_stability} hold for the iteration step $t$ from $0$ to $T$. The cost typically trades off between driving the voltages to the reference value and the control effort. 
The cost $C_v(\bm{v}(t)-\mathbbold{1}_n) $ typically be quantified as 2-norm, 1-norm or $\infty$-norm of the voltage deviation $\bm{v}(t)-\mathbbold{1}_n$~\cite{zhang2014optimal, vaccaro2011decentralized, jafari2018optimal}.
The control effort $C_q(\bm{q}(t))$ depends on the type of resources and can be both quadratic~\cite{zhao2014design, mallada2017optimal} and non-quadratic ones~\cite{shi2017using, vaccaro2011decentralized, jafari2018optimal}.  The proposed approach in this paper is applicable to all of these cost functions, and we do not make a distinction between them except for in the simulation section. 

The stability constraint in~\eqref{subeq:Optimization_stability} means that if we view \eqref{subeq:Optimization_q} and \eqref{subeq:Optimization_V} as a dynamical system with states $\bm{q}$ and $\bm{v}$, and input $\bm{u}$, then it is asymptotically stable as $t$ grows. Technically, if we only look at finite time horizons, the stability condition is not strictly necessary since all costs would be bounded and one could just optimize the cost. However, the initial states during online operation may differ from those in the optimization stage, and therefore a low optimization cost may not guarantee the stability of the dynamical system when starting from a new initial condition.
Therefore, we explicitly introduce the stability constraint and this leads to a constraint on the functional form of $\bm{u}$, which in turn improves the performance of the algorithm. 



In this paper, we aim to design a controller in $\bm{u}$ that explicitly responds to the time-varying power injections. We assume that each node $i$ can measure its local active power injection $p_i(t)$. Intuitively, we can design the control law $u_i(v_i,p_i)$ as a function of both voltage deviation and the present time net-load, which might enable the controller to cancel the impact of time variations in net power injections. The key challenge is that, even if the controller is implemented locally, it needs to stabilize the entire system, as characterized by the constraint \eqref{subeq:Optimization_stability}. In addition, voltage oscillations are influenced by future variations in power injections. Designing the controller solely based on active power measurements from the previous step cannot proactively address the influence of future power injections.

Instead of directly using the measurement of net power injection, we leverage the fact that although the net load is time-varying, it is largely predictable~\cite{xia2021stacked, wang2017data}. Embedding the predictions in the controllers may help the system adapt to disturbances brought by future variations in power injections. However, predictions contain errors. Then how can we embed the predictions in control, and how can we achieve performance guarantees with the prediction errors become the key questions to answer.
In the next sections, we will show our modeling of predictable parts using local measurable information and how we can achieve provable stabilizing guarantees by embedding the predictions in the design of adaptive controllers.

\section{Prediction Models}
\label{sec:prediction}

In this section, we first specify a prediction model. Then we derive the voltage dynamics under this model.  
\subsection{Time-varying power injections}
We model the time-varying net power injection as 
\begin{equation}\label{eq: p_t}
    p_{i}(t+1) = p_{i}(t) + \bm{c}_i^\top\bm{\phi}_{i}(t)+\Delta_{i}^p(t)\, ,
\end{equation}
where $\bm{c}_i^\top\bm{\phi}_{i}(t)+\Delta_{i}^p(t)$ captures differences of the power injections in successive time steps. More specifically, we think of $\bm{\phi}_{i}(t)$ as a vector of basis functions of features and $\bm{c}_i$ as the combination coefficients of the features (possibly unknown). Note that both $\bm{\phi}$ and $\bm{c}$ are indexed by bus $i$ and different buses may have a different set of basis functions and coefficients. The term $\Delta_{i}^p(t)$ represents errors that were not captured by $\bm{c}_i^\top\bm{\phi}_{i}(t)$.

The basis functions $\bm{\phi}_i$ can include features that are typically used in load forecasting algorithms, for example,  weather, temperature, time of the day, day of the week, solar irradiation, wind conditions, historical load and others~\cite{hong2020energy}. In addition, it can include different kernel functions that have been proposed in more recent forecasting algorithms~\cite{wu2019multiple,ghasempour2023electric}. It is important to note that we do not assume that $\bm{c}$ is known, that is, we are not reliant on a good forecasting algorithm. Rather, as long as the features are included in $\bm{\phi}_i$, our controller would adapt to them. This means that we can include a large set of features to fully capture the behavior of the net load.  



\subsection{Equivalent voltage dynamics }

To prepare for the decentralized controller design, we now derive equivalent transition dynamics by separating the local and nonlocal components in~\eqref{eq:Dynamic}. 
 Let $ \tilde{\bm{v}}(t)= \bm{v}(t)-\mathbbold{1}_n$ be the voltage difference from its reference value. 
From the LinDistFlow model in~\eqref{eq:Dyn_voltage}, we have
\begin{align}
 \tilde{\bm{v}}(t)&=\mathbf{R} \bm{p}(t)+\mathbf{X} \bm{q}(t).
\end{align}
Taking the difference between two consecutive time steps yields
\begin{align}\label{eq: vdiff}
 \tilde{\bm{v}}(t\!+\!1) \!- \!\tilde{\bm{v}}(t) &=\mathbf{R} \left(\bm{p}(t\!+\!1) \!-\! \bm{p}(t)\right)+\mathbf{X} \left(\bm{q}(t\!+\!1) - \bm{q}(t)\right).
\end{align}

Substituting controller \eqref{subeq:Dynamic_q} and predictor \eqref{eq: p_t} in~\eqref{eq: vdiff}, we have
\begin{equation}
\begin{split}
 \tilde{\bm{v}}(t\!+\!1) &= \tilde{\bm{v}}(t) + \mathbf{R} \left( \hat{\bm{c}}^\top\bm{\phi}(t)+\Delta^p(t)) \right)- \mathbf{X} \bm{u}(t) \\
  &= \tilde{\bm{v}}(t) \!-\! \mathbf{X}\left(\bm{u}(t)\!  - \! \mathbf{X}^{-1}\mathbf{R}(\hat{\bm{c}}^\top\bm{\phi}(t) \! +\! \Delta^p(t))\right),
\end{split}
\end{equation}
where $\bm{\phi}=\left(\bm{\phi}_1,\dots,\bm{\phi}_n\right)$ stacks basis functions and $\hat{\bm{c}}=\text{diag}(\bm{c}_1,\cdots,\bm{c}_n)$ stacks the coefficients $\bm{c}_i$ into a block diagonal matrix. Note that   $\hat{\bm{c}}^T \bm{\phi}(t)=\left(\bm{c}_1^\top\bm{\phi}_1(t),\cdots,\bm{c}_n^\top\bm{\phi}_n(t)\right)$, thus $\bm{p}(t\!+\!1) \!-\! \bm{p}(t)=\hat{\bm{c}}^T \bm{\phi}(t)+\Delta^p(t)$ . 

To separate the local and non-local terms, let $\mathbf{D}^e = \diag(\mathbf{X}^{-1} \mathbf{R})$ and $\mathbf{D}^o =\overline{\diag}(\mathbf{X}^{-1}\mathbf{R})$ be the diagnal and off-diagnal part of the matrix $\mathbf{X}^{-1} \mathbf{R}$, respectively.
Then
\begin{align}\label{eq:local-no-local}
 \tilde{\bm{v}}(t\!+\!1) &=\! \tilde{\bm{v}}(t) \!-\! \mathbf{X}\bm{u}(t) + \mathbf{R}\Delta^p(t)\\
 &\quad +
 \mathbf{X}\left( \mathbf{D}^e\hat{\bm{c}}^\top \bm{\phi}(t) +\mathbf{D}^o \hat{\bm{c}}^\top\bm{\phi}(t)\right) \nonumber
\end{align}
where $\mathbf{D}^e\hat{\bm{c}}^\top \bm{\phi}(t)$ is the term that only involves local information, while $\mathbf{D}^o \hat{\bm{c}}^\top\bm{\phi}(t)$ characterizes the impact of non-local time-varying injections.

Considering that the net load in different buses is typically correlated (for example, PV generation in a given region may exhibit similar output patterns),  local measurements can partially reflect the effects of time-varying net load at other buses. This allows our local control laws to (partially) account for the impact of $\mathbf{D}^o \hat{\bm{c}}^\top\bm{\phi}(t)$ to the extent of load correlations. To represent correlations in the features across the buses, let
$\bm{\phi}_{j}(t)=\Theta_{i,j}^{\text{cor}}\bm{\phi}_{i}(t)+\bm{\xi}^{\phi}_{i,j}(t)$ with $\Theta_{i,j}^{\text{cor}}$ being the correlation coefficient matrix, and $\bm{\xi}^{\phi}_{i,j}(t)$ being the vector for time variation of basis function in node $j$ that not correlate with the local basis functions in node $i$.\footnote{If all the features are independent, we can set $\Theta$ to be the zero matrix.}
Then, the non-local time-varying injections can be represented using the local basis function as
\begin{equation}\label{eq:corelate}
 (\mathbf{D}^o \hat{\bm{c}}^\top\bm{\phi}(t))_{i}=\bm{\theta}_{i}^{\text{cor}}\bm{\phi}_{i}(t)+\Delta^{\phi}_{i}(t),   
\end{equation}
where $\bm{\theta}_{i}^{\text{cor}}=\sum_{j\neq i}\mathbf{D}^o_{i,j} \bm{c}_{j}^\top\Theta_{i,j}^{\text{cor}}$ and $\Delta^{\phi}_i(t)=\sum_{j\neq i}\mathbf{D}^o_{[i,j]} \bm{c}_{j}^\top\bm{\xi}^{\phi}_{i,j}(t)$. Here $\bm{\theta}_{i}^{\text{cor}}\bm{\phi}_{i}(t)$ is the time-variation of the non-local net load that can be captured by local basis functions, and $\Delta^{\phi}_i(t)$ is the residual part.


Substituting~\eqref{eq:corelate} into~\eqref{eq:local-no-local}, the voltage dynamics under time-varying netload are written as
\begin{equation}\label{eq: dyn_voltage}
\tilde{\bm{v}}(t\!+\!1)=\tilde{\bm{v}}(t)-\bm{X}\left(\bm{u}(t)-\hat{\bm{\phi}}(t)^\top \bm{a}\right)+\bm{\delta}^{v}(t),
\end{equation}
where $\bm{a}:=(\bm{a}_1,\cdots, \bm{a}_n)$ and $\bm{a}_i :=\mathbf{D}^e_{ii}\bm{c}_i+\bm{\theta}_{i}^{\text{cor}}$ for $i\in [n]$.
The term $\hat{\bm{\phi}}(t)$ is constructed by  stacking locally defined basis $\bm{\phi}_{i}(t)$ diagonally, written as $\hat{\bm{\phi}}(t)=\text{diag}\left(\bm{\phi}_{1}(t),\cdots,\bm{\phi}_{n}(t)\right)$.
The lumped residual term is $\bm{\delta}^{v}(t) = \mathbf{X}\Delta^{\phi}_i(t)+\mathbf{R}\Delta^p(t)$, where the first term is the residual from non-local real power injection and the second term is the prediction error.

Intuitively,  if the controller is of the form $u_{i}(t) = k_i\hat{v}_{i}(t)+\bm{\phi}_{i}(t)^\top\bm{a}_i$, it can cancel the impact of net load variations in $\bm{\phi}_{i}(t)^\top\bm{a}_i$.
However, this is not possible since the coefficients $\bm{a}_i$ is not known, and thus $\bm{\phi}_{i}(t)^\top\bm{a}_i$ is not directly computable. Following the work in ~\cite{slotine1991applied,o2022neural,shi2021meta}, we design an adaptation law $\bm{\phi}_{i}(t)^\top\bm{\tilde{a}}_{i}(t)$, and the next section describes how it is updated.

\section{Modular Design of Adaptive Controllers } \label{sec:module}


In this section, we propose an adaptive approach to leverage local predictions in voltage control. On this basis, we derive the equilibrium of the closed-loop system and conditions for convergence towards the equilibrium. 
\subsection{Adaptive control law}
 
We adopt a modular approach for the controller design to enable adaptation to time-varying net load with minimal modifications to existing voltage control laws. Specifically, we design the control law as the summation of two parts: 1) Base Controller: linear control  that aligns with the IEEE 1547 standard for voltage regulation~\cite{photovoltaics2018ieee}, which works well
for systems with a time-invariant net load; 2) Adaptation Law, the control law $\bm{\phi}_{i}(t)^\top\bm{\tilde{a}}_{i}(t)$ 
 for the time-varying load, with $\bm{\tilde{a}}_{i}$ being an additional state serving as the adaptation coefficient. Compactly,  the control law is
\begin{subequations}\label{eq:control_adpt}
\begin{align}
& u_{i}(t) = \underbrace{k_i\tilde{v}_{i}(t)}_{\text{Base controller}}+\;\underbrace{\bm{\phi}_{i}(t)^\top\bm{\tilde{a}}_{i}(t)}_{\text{Adaptive Law}}\\
&\tilde{\bm{a}}_{i}(t\!+\!1)=\alpha \tilde{\bm{a}}_{i}(t)+\tilde{v}_{i}(t)\cdot\mathbf{A}_i\bm{\phi}_{i}(t),\label{subeq: dot_hat_a}
\end{align}
\end{subequations}
where $k_i$ is the linear control gain, $\mathbf{A}_i\succ 0$ is a tunable matrix, and $0<\alpha<1$ is a tunable scalar.

Since the closed-loop system~\eqref{eq: dyn_voltage} and~\eqref{eq:control_adpt} is affected by time-varying basis function $\bm{\phi}(t)$ and the error $\bm{\delta}(t)$, the system does not have a time-invariant equilibrium. In the next section, we first show that the system have a well-defined time-varying equilibrium if choosing $\alpha$ properly, and the voltage deviation at the equilibrium can be close to zero. Next, we derive conditions to achieve the exponential input-to-state stability of the system around the equilibrium.

\subsection{Equilibrium}
The closed-loop system formed by~\eqref{eq: dyn_voltage} and~\eqref{eq:control_adpt} can be written as
\begin{equation}\label{eq:closed-loop}
    \begin{split}
      \tilde{\bm{v}}(t+1)&=\!\tilde{\bm{v}}(t)\!-\! \mathbf{X}\!\left( \!\hat{\mathbf{K}} \tilde{\bm{v}}(t)\!-\! \hat{\bm{\phi}}(t)^{\top}\!\!\left(\bm{a}\!-\!\tilde{\bm{a}}(t)\!\right)\!\right)+\bm{\delta}^{v}(t), \\   
            \tilde{\bm{a}}(t+1)&=\alpha \tilde{\bm{a}}(t)+\hat{\mathbf{A}}\hat{\bm{\phi}}(t) \tilde{\bm{v}}(t),  
    \end{split}
\end{equation}
where $\hat{\mathbf{A}}:=\text{diag}(\mathbf{A}_1\cdots,\mathbf{A}_n)$ and $\hat{\mathbf{K}}:=\text{diag}(k_1\cdots,k_n).$

The equilibrium of this closed-loop system is characterized by the following lemma.
\begin{lem}\label{lem: equilibrium}
    The equilibrium $(\tilde{\bm{v}}^*(t), \tilde{\bm{a}}^*(t))$ of the closed-loop system~\eqref{eq: dyn_voltage} and~\eqref{eq:control_adpt} at the time $t$ satisfies $\tilde{\bm{v}}^*(t)=\left(\hat{\mathbf{K}}+\frac{1}{1-\alpha} \hat{\bm{\phi}}(t)^{\top} \hat{\mathbf{A}}\hat{\bm{\phi}}(t)\right)^{-1} \!\!\left(\! \hat{\bm{\phi}}(t)^{\!\top} \bm{a}\!+\!\mathbf{X}^{\!-1}\bm{\delta}^{v}(t)\!\right) $ and $ \tilde{\bm{a}}^*(t)=\frac{1}{1-\alpha} \hat{\mathbf{A}} \hat{\bm{\phi}}(t) \bm{v}_t^{*}$. 
    
    When $\alpha \rightarrow 1$, the equilibrium approaches   
1) $\tilde{\bm{v}}^*(t) \rightarrow \mathbbold{0}_n $,
and 2) $\hat{\bm{\phi}}(t)^{\top} \tilde{\bm{a}}^*(t) \rightarrow \hat{\bm{\phi}}(t)^{\top} \bm{a}+\mathbf{X}^{\!-1}\bm{\delta}^{v}(t)$.
\end{lem}
Note the dependence on $t$ of $\tilde{\bm{v}}^{*}(t)$ and hence it is useful to think of this equilibrium as a (slowly) time-varying target, and the trajectories are controlled to track this target. This time dependence cannot be completely eliminated because of the prediction error in the load and nonlocal influences.

On the other hand, as shown in Lemma~\ref{lem: equilibrium},
when $\alpha$ approaches 1, the equilibrium approaches $\tilde{\bm{v}}^{*}(t)=0$ (that is, the voltages are at the reference value). However, we also need $\alpha<1$ to ensure that the impact of the error term $\bm{\delta}^v(t)$ will decay through time.
Later in Theorem~\ref{thm:iss}, we will show that $0<\alpha<1$ guarantees the convergence of states and also input-to-state stable with respect to error terms. Therefore, we set $\alpha$ to be strictly less than but close to 1 ( empirically about 0.99) to achieve $\tilde{\bm{v}}^*(t) \approx  \mathbbold{0}_n $ and ensure desirable convergence performance.    





The proof of Lemma~\ref{lem: equilibrium} is given below.
\begin{proof}
Setting the left side of equations ~\eqref{eq:closed-loop} equals to the right side gives the  equilibrium 
\begin{subequations}
    \begin{align}
           \mathbbold{0}_n&=- \mathbf{X} \hat{\mathbf{K}} \tilde{\bm{v}}^*(t)+ \mathbf{X} \hat{\bm{\phi}}(t)^{\top}\left(\bm{a}-\tilde{\bm{a}}^*(t)\right)+\bm{\delta}^{v}(t)\label{subeq:equlibrium-v} \\   
            \tilde{\bm{a}}^*(t)&=\alpha \tilde{\bm{a}}^*(t)+\hat{\mathbf{A}}\hat{\bm{\phi}}(t) \tilde{\bm{v}}^*(t).\label{subeq:equlibrium-a}
    \end{align}
\end{subequations}


Since $\mathbf{X}$ is invertible ($\mathbf{X}\succ0$), \eqref{subeq:equlibrium-v} is equivalent to $\hat{\mathbf{K}} \tilde{\bm{v}}^*(t)=\hat{\bm{\phi}}(t)^{\top}\left(\bm{a}-\tilde{\bm{a}}^*(t)\right) +\mathbf{X}^{-1}\bm{\delta}^{v}(t)$. Further plugging in \eqref{subeq:equlibrium-a} yields
\begin{subequations}\label{eq:equ}
    \begin{align}
        \tilde{\bm{v}}^*(t)&=\left(\!\hat{\mathbf{K}}\!+\!\frac{1}{1\!-\!\alpha} \hat{\bm{\phi}}(t)^{\!\top} \!\hat{\mathbf{A}}\hat{\bm{\phi}}(t)\!\right)^{\!\!-1}\!\!\left(\! \hat{\bm{\phi}}(t)^{\!\top} \bm{a}\!+\!\mathbf{X}^{\!-1}\bm{\delta}^{v}(t)\!\right)\label{subeq:v*}\\
       \tilde{\bm{a}}^*(t)&=\frac{1}{1-\alpha} \hat{\mathbf{A}} \hat{\bm{\phi}}(t) \tilde{\bm{v}}_t^{*}\label{subeq:a*}
    \end{align}
\end{subequations}

When $\alpha \rightarrow 1$, $\frac{1}{1-\alpha} \hat{\bm{\phi}}_i(t)^{\top} \hat{\mathbf{A}}_i\hat{\bm{\phi}}_i(t)\rightarrow \infty$. Thus $\tilde{\bm{v}}^*(t) \rightarrow 0$ from the relation in~\eqref{subeq:v*}.

From~\eqref{subeq:v*} and~\eqref{subeq:a*}, we have
\begin{equation}
\begin{split}
    &\hat{\bm{\phi}}(t)^{\top}\tilde{\bm{a}}^*(t)\\
    &=\frac{1}{1-\alpha} \hat{\bm{\phi}}(t)^{\top}\hat{\mathbf{A}}\hat{\bm{\phi}}(t) \left(\hat{\mathbf{K}}+\frac{1}{1-\alpha} \hat{\bm{\phi}}(t)^{\top} \hat{\mathbf{A}}\hat{\bm{\phi}}(t)\right)^{-1}\\
    &\qquad\cdot\left(\! \hat{\bm{\phi}}(t)^{\!\top} \bm{a}\!+\!\mathbf{X}^{\!-1}\bm{\delta}^{v}(t)\!\right)\\
    & \overset{\underset{\mathrm{\alpha \rightarrow 1}}{}}{=}\frac{1}{1-\alpha} \hat{\bm{\phi}}(t)^{\top}\hat{\mathbf{A}}\hat{\bm{\phi}}(t) \left(\frac{1}{1-\alpha} \hat{\bm{\phi}}(t)^{\top} \hat{\mathbf{A}}\hat{\bm{\phi}}(t)\right)^{-1}\\
    &\qquad\cdot\left(\! \hat{\bm{\phi}}(t)^{\!\top} \bm{a}\!+\!\mathbf{X}^{\!-1}\bm{\delta}^{v}(t)\!\right)\\
    &=\hat{\bm{\phi}}(t)^{\top} \bm{a}+\mathbf{X}^{\!-1}\bm{\delta}^{v}(t)
\end{split}
\end{equation}
Therefore, when $\alpha \rightarrow 1$, the equilibrium approaches   
1) $\tilde{\bm{v}}^*(t) \rightarrow \mathbbold{0}_n $,
and 2) $\hat{\bm{\phi}}(t)^{\top} \tilde{\bm{a}}^*(t) \rightarrow \hat{\bm{\phi}}(t)^{\top} \bm{a}+\mathbf{X}^{\!-1}\bm{\delta}^{v}(t)$.
\end{proof}

\subsection{State transition dynamics}
Using the representation of the equilibrium in~\eqref{eq:equ}, we can eliminate terms in ~\eqref{eq:closed-loop} and express the closed-loop system~\eqref{eq:closed-loop} equivalently as
\begin{equation}\label{eq:closed-partial}
    \begin{split}
       &\begin{bmatrix}
\tilde{\bm{v}}(t\!+\!1)-\tilde{\bm{v}}^*(t) \\
\hat{\bm{\phi}}(t)^{\top}\left(\tilde{\bm{a}}(t\!+\!1)-\tilde{\bm{a}}^*(t)\right)
\end{bmatrix}\\
&=\begin{bmatrix}
\mathbf{I}_n- \mathbf{X}\hat{\mathbf{K}} & - \mathbf{X} \\
\hat{\bm{\phi}}(t)^{\top} \hat{\mathbf{A}} \hat{\bm{\phi}}(t) & \alpha \mathbf{I}_n
\end{bmatrix}\begin{bmatrix}
\tilde{\bm{v}}(t)-\tilde{\bm{v}}^*(t) \\
\hat{\bm{\phi}}(t)^{\top}\left(\tilde{\bm{a}}(t)-\tilde{\bm{a}}^*(t)\right)
\end{bmatrix}
    \end{split}
\end{equation}


To analyze the evolution of states through time, we need the recursion between $\begin{bmatrix}
\tilde{\bm{v}}(t\!+\!1)-\tilde{\bm{v}}^*(t+1) \\
\hat{\bm{\phi}}(t+1)^{\top}\left(\tilde{\bm{a}}(t\!+\!1)-\tilde{\bm{a}}^*(t+1)\right)
\end{bmatrix}$ and $\begin{bmatrix}
\tilde{\bm{v}}(t)-\tilde{\bm{v}}^*(t) \\
\hat{\bm{\phi}}(t)^{\top}\left(\tilde{\bm{a}}(t)-\tilde{\bm{a}}^*(t)\right)
\end{bmatrix}$. To this end, we represent the left side of~\eqref{eq:closed-partial} as 
\begin{equation}
    \begin{split}
       & \begin{bmatrix}
        \tilde{\bm{v}}(t\!+\!1)-\tilde{\bm{v}}^*(t) \\
        \hat{\bm{\phi}}(t)^{\top}\left(\tilde{\bm{a}}(t\!+\!1)-\tilde{\bm{a}}^*(t)\right)
        \end{bmatrix}\\
        &= \begin{bmatrix}
        \tilde{\bm{v}}(t\!+\!1)-\tilde{\bm{v}}^*(t+1) \\
        \hat{\bm{\phi}}(t+1)^{\top}\left(\tilde{\bm{a}}(t\!+\!1)-\tilde{\bm{a}}^*(t+1)\right)
        \end{bmatrix}
        +\begin{bmatrix}
        \bm{\rho}^v(t) \\
        \bm{\rho}^a(t),
        \end{bmatrix}
\end{split}
\end{equation}
where $\bm{\rho}^v(t)=\tilde{\bm{v}}^*(t+1)-\tilde{\bm{v}}^*(t) $ and $\bm{\rho}^a(t)=\hat{\bm{\phi}}(t)^{\top}\left(\tilde{\bm{a}}(t\!+\!1)-\tilde{\bm{a}}^*(t)\right)-\hat{\bm{\phi}}(t+1)^{\top}\left(\tilde{\bm{a}}(t\!+\!1)-\tilde{\bm{a}}^*(t+1)\right)$.

Denote the transition matrix as $$\mathbf{M}(t):=\begin{bmatrix}
\mathbf{I}_n- \mathbf{X}\hat{\mathbf{K}} & - \mathbf{X} \\
\hat{\bm{\phi}}(t)^{\top} \hat{\mathbf{A}} \hat{\bm{\phi}}(t) & \alpha \mathbf{I}_n
\end{bmatrix}.$$
The closed-loop system formed by~\eqref{eq: dyn_voltage} and~\eqref{eq:control_adpt} can now be written as
\begin{equation}\label{eq:transition-rec}
    \begin{split}
       &\begin{bmatrix}
\tilde{\bm{v}}(t\!+\!1)-\tilde{\bm{v}}^*(t\!+\!1) \\
\hat{\bm{\phi}}(t+1)^{\top}\left(\tilde{\bm{a}}(t\!+\!1)-\tilde{\bm{a}}^*(t+1)\right)
\end{bmatrix}\\
&=\mathbf{M}(t)\begin{bmatrix}
\tilde{\bm{v}}(t)-\tilde{\bm{v}}^*(t) \\
\hat{\bm{\phi}}(t)^{\top}\left(\tilde{\bm{a}}(t)-\tilde{\bm{a}}^*(t)\right)
\end{bmatrix}
+\begin{bmatrix}
\bm{\rho}^v(t) \\
\bm{\rho}^a(t)
\end{bmatrix}
    \end{split}
\end{equation}


Thus, the convergence of the states in $\left(\tilde{\bm{v}(t)}, \tilde{\bm{a}}(t)\right)$ is characterized by the eigenvalues of the transition matrix $\mathbf{M}(t)$. In the next section, we show conditions to bound the eigenvalues of $\mathbf{M}(t)$ and how this leads to input-to-state stability of the closed-loop system. 


\subsection{Input-to-State Stability}
We characterize the convergence of states in the following theorem.
\begin{thm}\label{thm:iss}
    If  the  eigenvalues
    of the transition matrix satisfy $\max_j|\lambda_j\left(\mathbf{M}(t)\right)|\leq 1-\epsilon$ for a small positive constant $\epsilon\in(0,1)$,
    then  the state deviation is bounded by $ \left\| \begin{bmatrix}
\tilde{\bm{v}}(t)-\tilde{\bm{v}}^*(t) \\
\hat{\bm{\phi}}(t)^{\top}\left(\tilde{\bm{a}}(t)-\tilde{\bm{a}}^*(t)\right)
\end{bmatrix}\right\|_2\leq (1-\epsilon)^t \left\|\begin{bmatrix}
\tilde{\bm{v}}(0)-\tilde{\bm{v}}^*(0) \\
\hat{\bm{\phi}}(0)^{\top}\left(\tilde{\bm{a}}(0)-\tilde{\bm{a}}^*(0)\right)
\end{bmatrix}\right\|_2+\frac{1-\epsilon^{t}}{1-\epsilon}\Bar{\bm{\rho}}$, where $\Bar{\bm{\rho}}=\max_t\sqrt{\left\| \bm{\rho}^v(t)\right\|_2^2+\left\| \bm{\rho}^a(t)\right\|_2^2}$. Namely, the closed-loop system formed by~\eqref{eq: dyn_voltage} and~\eqref{eq:control_adpt} is input-to-state stable. 
\end{thm}
The proof is a variant of that given in Example 3.4 of~\cite{jiang2001input}. 
\begin{proof}
 Denote $\bm{x}(t)=\begin{bmatrix}
\tilde{\bm{v}}(t)-\tilde{\bm{v}}^*(t) \\
\hat{\bm{\phi}}(t)^{\top}\left(\tilde{\bm{a}}(t)-\tilde{\bm{a}}^*(t)\right)
\end{bmatrix}$, then the transition dynamics in~\eqref{eq:transition-rec} is written as
\begin{equation}\label{eq: transition-ab}
\bm{x}(t+1) = \bm{M}(t)\bm{x}(t) + \begin{bmatrix}
\bm{\rho}^v(t) \\
\bm{\rho}^a(t)
\end{bmatrix}
\end{equation}
where  the eigenvalues of $M(t)$ are at most $1-\epsilon$. 
Expanding the transition dynamics from the time $0$ to $t$ yields
\begin{equation}
x(t) = \prod_{j=0}^{t-1} M(j)x(0) + \begin{bmatrix}
\bm{\rho}^v(t\!-\!1) \\
\bm{\rho}^a(t\!-\!1)
\end{bmatrix}+\sum_{k=0}^{t-2} \prod_{j=k+1}^{t-1} \!\!M(j)\begin{bmatrix}
\bm{\rho}^v(k) \\
\bm{\rho}^a(k)
\end{bmatrix}, 
\end{equation}

Therefore,
\begin{equation}
\begin{split}
\left\|x(t) \right\|_2
& \leq (1-\epsilon)^t \left\|x(0)\right\|_2+\sum_{k=0}^{t-1} (1-\epsilon)^{t-1-k} \left\| \begin{bmatrix}
\bm{\rho}^v(k) \\
\bm{\rho}^a(k)
\end{bmatrix}\right\|_2\\
&\leq (1-\epsilon)^t \left\|x(0)\right\|_2+\left(\sum_{k=0}^{t-1} (1-\epsilon)^{t-1-k} \right)\Bar{\bm{\rho}}\\
&= (1-\epsilon)^t \left\|x(0)\right\|_2+\frac{1-\epsilon^{t}}{1-\epsilon}\Bar{\bm{\rho}}
\end{split}
\end{equation}

\end{proof}

Theorem~\ref{thm:iss} shows that the system is input-to-state stable if the eigenvalue of the transition matrix is no larger than $1-\epsilon$.  
The next theorem derives a sufficient condition to bound the eigenvalue of $\mathbf{M}(t)$.

\begin{thm}\label{lem: stability bounds}
The eigenvalues of  $\mathbf{M}(t):=\begin{bmatrix}
\mathbf{I}_n- \mathbf{X}\hat{\mathbf{K}} & - \mathbf{X} \\
\hat{\bm{\phi}}(t)^{\top} \hat{\mathbf{A}} \hat{\bm{\phi}}(t) & \alpha \mathbf{I}_n
\end{bmatrix}$ satisfy $\max_j|\lambda_j\left(\mathbf{M}(t)\right)|\leq 1-\epsilon$ for a small positive constant $\epsilon\in(0,1)$ if the following conditions hold:
    \begin{enumerate}[label=(\alph*)]
        \item $-(1-\epsilon)\mathbf{I}_n \preceq \mathbf{I}_n- \mathbf{X}^{\frac{1}{2}}\hat{\mathbf{K}}  \mathbf{X}^{\frac{1}{2}} \preceq (1-\epsilon)\mathbf{I}_n $
        \item $0<\alpha\leq1-\epsilon$
        \item $\lambda_{max}(\mathbf{X}^{\frac{1}{2}} \hat{\bm{\phi}}(t)^{\top} \hat{\mathbf{A}} \hat{\bm{\phi}}(t) \mathbf{X}^{\frac{1}{2}} )+\alpha\lambda_{max}( \mathbf{I}_n- \mathbf{X}^{\frac{1}{2}}\hat{\mathbf{K}}  \mathbf{X}^{\frac{1}{2}})\leq1-\epsilon$
    \end{enumerate}
\end{thm}

Notably, condition (a) in Theorem~\ref{lem: stability bounds} is equivalent to the exponential stability of time-invariant
system as mentioned in previous literatures~\cite{cui2022decentralized}. We elaborate this equivalence in Remark~\ref{rmk:K}. Condition (b) holds by setting $\epsilon$ a positive and small real number. Condition (c) then holds when $\bm{A}$ is small enough. Therefore, these conditions are not difficult to satisfy. The proof of  Theorem~\ref{lem: stability bounds} is given below.

\begin{proof}
    Let $(\bm{y}_1, \bm{y}_2)$ be the eigenvector corresponding to the transition matrix
    
$$
\begin{bmatrix}
\mathbf{I}_n- \mathbf{X}\hat{\mathbf{K}} & - \mathbf{X} \\
\hat{\bm{\phi}}(t)^{\top} \hat{\mathbf{A}} \hat{\bm{\phi}}(t) & \alpha \mathbf{I}_n
\end{bmatrix}
\begin{bmatrix}
\bm{y}_1 \\
\bm{y}_2
\end{bmatrix}
=\lambda \left[\begin{array}{cc}
\bm{y}_1 \\
\bm{y}_2
\end{array}\right]
$$
Then we have 
\begin{subequations}\label{eq: def_eigen}
\begin{equation}
         (\mathbf{I}_n- \mathbf{X}\hat{\mathbf{K}})\bm{y}_1- \mathbf{X}\bm{y}_2=\lambda \bm{y}_1,
\end{equation} 
\begin{equation}
   \hat{\bm{\phi}}(t)^{\top} \hat{\mathbf{A}} \hat{\bm{\phi}}(t)  \bm{y}_1+\alpha \bm{y}_2=\lambda \bm{y}_2.  
\end{equation}
\end{subequations}

If $\alpha=\lambda$, then 
$0<\lambda\leq1-\epsilon$ from the condition (b) in Lemma~\ref{lem: stability bounds}.

Therefore, the rest of the proof analyze the case when  $\alpha\neq\lambda$.

Eliminating $\bm{y}_2$ in~\eqref{eq: def_eigen} yields
$$\left((\mathbf{I}_n- \mathbf{X}\hat{\mathbf{K}})-\frac{1}{\lambda-\alpha} \mathbf{X}\hat{\bm{\phi}}(t)^{\top} \hat{\mathbf{A}} \hat{\bm{\phi}}(t) \right)\bm{y}_1=\lambda \bm{y}_1.
$$

Equivalently, we  have
\begin{equation*}
    \begin{split}
     &\mathbf{X}^{\frac{1}{2}}\!\left(\!(\mathbf{I}_n\!-\! \mathbf{X}^{\frac{1}{2}} \hat{\mathbf{K}}  \mathbf{X}^{\frac{1}{2}}\!)
    \!-\!\frac{1}{\lambda\!-\!\alpha}( \mathbf{X}^{\frac{1}{2}} \hat{\bm{\phi}}(t)^{\top} \!\hat{\mathbf{A}} \hat{\bm{\phi}}(t) \! \mathbf{X}^{\frac{1}{2}}\!)\!\right)\!\mathbf{X}^{-\frac{1}{2}}\bm{y}_1\\
    &=\lambda  \mathbf{X}^{\frac{1}{2}} \mathbf{X}^{-\frac{1}{2}} \bm{y}_1.    
    \end{split}
\end{equation*}

Since $\mathbf{X}\succ 0$, then we have
\begin{equation*}
    \begin{split}
     &\left(\!(\mathbf{I}_n\!-\! \mathbf{X}^{\frac{1}{2}} \hat{\mathbf{K}}  \mathbf{X}^{\frac{1}{2}}\!)
    \!-\!\frac{1}{\lambda\!-\!\alpha}( \mathbf{X}^{\frac{1}{2}} \hat{\bm{\phi}}(t)^{\top} \!\hat{\mathbf{A}} \hat{\bm{\phi}}(t) \! \mathbf{X}^{\frac{1}{2}}\!)\!\right)\!\mathbf{X}^{-\frac{1}{2}}\bm{y}_1\\
    &=\lambda  \mathbf{X}^{-\frac{1}{2}} \bm{y}_1,   
    \end{split}
\end{equation*}
which indicates that $(\mathbf{I}_n- \mathbf{X}^{\frac{1}{2}} \hat{\mathbf{K}} \mathbf{X}^{\frac{1}{2}})
-\frac{1}{\lambda-\alpha}( \mathbf{X}^{\frac{1}{2}} \hat{\bm{\phi}}(t)^{\top} \hat{\mathbf{A}} \hat{\bm{\phi}}(t)  \mathbf{X}^{\frac{1}{2}})-\lambda \mathbf{I}_n$ loose rank.

Therefore, there exists $\bm{z}\in\real^n, \bm{z}^\top \bm{z}=1$ such that
\begin{equation}\label{eq: eigen_extend}
\begin{split}
    &\bm{z}^\top(\mathbf{I}_n- \mathbf{X}^{\frac{1}{2}} \hat{\mathbf{K}} \mathbf{X}^{\frac{1}{2}})\bm{z}
    -\frac{1}{\lambda-\alpha}\bm{z}^\top( \mathbf{X}^{\frac{1}{2}} \hat{\bm{\phi}}(t)^{\top} \hat{\mathbf{A}} \hat{\bm{\phi}}(t)  \mathbf{X}^{\frac{1}{2}})\bm{z} \\
    & -\lambda \bm{z}^\top \bm{z}=0.   
\end{split}
\end{equation}
Denote $\mu_K := \bm{z}^\top(\mathbf{I}_n- \mathbf{X}^{\frac{1}{2}} \hat{\mathbf{K}} \mathbf{X}^{\frac{1}{2}})\bm{z}$, 
$\mu_\phi := \bm{z}^\top( \mathbf{X}^{\frac{1}{2}} \hat{\bm{\phi}}(t)^{\top} \hat{\mathbf{A}} \hat{\bm{\phi}}(t)  \mathbf{X}^{\frac{1}{2}})\bm{z}$. Then~\eqref{eq: eigen_extend}  is equivalent to
\begin{equation}\label{eq: eigen_extend2}
\mu_K
-\frac{1}{\lambda-\alpha}\mu_\phi-\lambda =0, 
\end{equation}
where the condition $ -(1-\epsilon)\mathbf{I}_n \preceq \mathbf{I}_n- \mathbf{X}^{\frac{1}{2}}\hat{\mathbf{K}} \mathbf{X}^{\frac{1}{2}} \preceq (1-\epsilon)\mathbf{I}_n $ ensures that $-(1-\epsilon)\leq\mu_K\leq(1-\epsilon)$.

Note that $\lambda-\alpha\neq0$, then multiplying both sides of~\eqref{eq: eigen_extend2} with  $\lambda-\alpha$ yields
\begin{equation}\label{eq: eigen_extend3}
\left(\lambda-\frac{1}{2}(\mu_K+\alpha)\right)^2= \frac{1}{4}(\mu_K-\alpha)^2-\mu_\phi.  
\end{equation}

\begin{enumerate}
    \item If $\frac{1}{4}(\mu_K-\alpha)^2-\mu_\phi>0$, then $\lambda$ is real.

    The solution of~\eqref{eq: eigen_extend2} can be explicitly written as
    \begin{equation}
        \lambda_{+,-}=\frac{1}{2}(\mu_K+\alpha)\pm \sqrt{\frac{1}{4}(\mu_K-\alpha)^2-\mu_\phi},
    \end{equation}
    Thus 
   \begin{equation}
   \begin{split}
    \lambda_{+}&<\frac{1}{2}(\mu_K+\alpha)+\frac{1}{2}|\mu_K-\alpha|\\
    &<\max(\mu_K, \alpha)\\
    &\leq(1-\epsilon),   
   \end{split}
   \end{equation}
   and
   \begin{equation}
   \begin{split}
       \lambda_{-}&>\frac{1}{2}(\mu_K+\alpha)-\frac{1}{2}|\mu_K-\alpha|\\
        &>\min(\mu_K, \alpha)\\
        &\geq-(1-\epsilon),
   \end{split}
   \end{equation}
    
    \item If $\frac{1}{4}(\mu_K-\alpha)^2-\mu_\phi<0$, then $\lambda$ is complex.

    In this case, the solution of~\eqref{eq: eigen_extend2} can be explicitly written as
    \begin{equation}
        \lambda_{+,-}=\frac{1}{2}(\mu_K+\alpha)\pm j\sqrt{-\frac{1}{4}(\mu_K-\alpha)^2+\mu_\phi},
    \end{equation}

    Thus 
   \begin{equation}
   \begin{split}
        |\lambda|^2&=\left(\frac{1}{4}(\mu_K+\alpha)^2-\frac{1}{4}(\mu_K-\alpha)^2+\mu_\phi\right)\\
        &=\mu_K\alpha+\mu_\phi\\
        &\leq1-\epsilon,
   \end{split}
   \end{equation}
    where the last inequality follows condition (c) in Lemma~\ref{lem: stability bounds}.

\end{enumerate}

\end{proof}

\begin{rmk}[equivalent condition for $\mathbf{I}_n- \mathbf{X}^{\frac{1}{2}}\hat{\mathbf{K}} \mathbf{X}^{\frac{1}{2}} $]\label{rmk:K}
    The condition  $-(1-\epsilon)\mathbf{I}_n \preceq \mathbf{I}_n- \mathbf{X}^{\frac{1}{2}}\hat{\mathbf{K}} \mathbf{X}^{\frac{1}{2}} \preceq (1-\epsilon)\mathbf{I}_n $ guarantees the exponential stability of time-invariant system. This condition holds if $\epsilon\mathbf{X}^{-1} \preceq \hat{\mathbf{K}} \preceq (2-\epsilon)\mathbf{X}^{-1} $, which can be enforced decentralizedly through an estimation of eigenvalues of $\mathbf{X}^{-1}$.
\end{rmk}
\begin{rmk}[Tuning of $\epsilon$]\label{rmk:epsilon}
         Note that a smaller $1-\epsilon$ leads to a smaller real part of the eigenvalue of the transition matrix, which leads to faster convergence to the equilibrium. However, since the decay rate of states in Theorem~\ref{thm:iss} is $(1-\epsilon)^t$, the state deviations can still decay fast enough even if $1-\epsilon$ is close to one. Therefore, we still recommend set up $\epsilon$ to be a small positive scalar such that $\alpha$ is close to one and $\tilde{\bm{v}}^*(t) \approx  \mathbbold{0}_n $.
\end{rmk}

We further derive the following decentralized stability condition that can be enforced conveniently without extra coordination between nodes.
\begin{corollary}[Decentralized stability conditions] \label{cor:1}
The conditions in Theorem~\ref{lem: stability bounds} can be satisfied decentralized by
\begin{enumerate}[label=(\alph*)]
    \item $\epsilon \lambda_{\max}(\mathbf{X}^{-1} )\leq k_i\leq (2-\epsilon)\lambda_{\min}(\mathbf{X}^{-1})$
    \item $0<\alpha \leq 1-\epsilon$
    \item $\phi_i^\top\bm{A}_i\phi_i\leq (1-\epsilon)(1-\alpha)/\lambda_{max}(\mathbf{X)}$
\end{enumerate}
\end{corollary}
The conditions in Corollary~\ref{cor:1} is more conservative than those found in Theorem~\ref{thm:iss}. In particular, depending on the eigenvalues of $\bd{X}$,  $\epsilon$ may need to be small, thus leading to slower convergence (although still exponential in rate). Nevertheless, if communication between nodes are expensive, it provides a decentralized adaptation algorithm.

\subsection{Parameter tuning and optimizations}

Theorem~\ref{lem: stability bounds} provides the conditions on the control parameters to guarantee the stability of the closed-loop systems. The transient performance of the system (i.e., the dynamic performance of the system before reaching the equilibrium), is affected largely by how the parameters are tuned.

In this work, we assume that the 
matrix $\mathbf{X}$ is available. This assumption is justified by the fact that $\mathbf{X}$ (and $\mathbf{R}$) can be inferred from historical smart meter data collected over extended periods~\cite{yu2018patopaem,zhang2020topology,lin2021data}, where data reporting rates can be relatively infrequent (e.g., once per day~\cite{kavetcharacterization}). Note that the controllers in~\eqref{eq:control_adpt} only depend on the local information. Therefore, the implementation of the controller in the real time does not require communication capabilities, while the offline tuning and optimization could be done with information $\mathbf{X}$ and $\mathbf{R}$ in a centralized manner.

The controllers are parameterized through~\eqref{eq:control_adpt} and satisfying conditions (a)-(c) in Theorem~\ref{lem: stability bounds} by design. The tunable parameters include $\epsilon, \alpha$ and $\mathbf{A}_i, k_i$ for each node $i$. Note that we hope that $\alpha$ is close to one for an equilibrium where $\tilde{\bm{v}}^* \approx 0$. Therefore, we set up $\epsilon$ to be a small constant (equals to 0.01 in our experiment) and $\alpha=1-\epsilon$. The parameters $\mathbf{A}_i, k_i$ are optimized by training through the training framework given in Algorithm 1.  At each epoch of training, we collect a batch of trajectories under the current controller. Let $H$ be the number of batches, and the trajectory in each batch be $\left\{\bm{v}_h(1),\bm{q}_h(1),\cdots,\bm{v}_h(T),\bm{q}_h(T)\right\}$. The parameters $\bm{A}$ and $\bm{k}$ are then optimized through the recurrent neural network-based training methods in~\cite{cui2020}, which conduct gradient descent using the following loss function
\begin{equation}
    Loss=\quad \sum_{h=1}^{H}\sum_{t=1}^{T} C_q(\bm{q}_h(t))+C_v(\bm{v}_h(t)-\mathbbold{1}_n), 
\end{equation}
where the cost functions $C_q(\cdot)$ and $C_v(\cdot)$ are defined the same as~\eqref{eq:Optimization}.

\begin{algorithm}
 \caption{ Reinforcement Learning for Controller Parameter Optimization~\cite{cui2020} }
 \begin{algorithmic}[1]
 \renewcommand{\algorithmicrequire}{\textbf{Require: }}
 \renewcommand{\algorithmicensure}{\textbf{Input:}}
 \REQUIRE  Learning rate $\alpha$, batch size $H$, total time steps T, number of epochs $I$, paramter $\epsilon$ in the controller~\eqref{eq:control_adpt}
 \ENSURE The initial voltage $\bm{v}(0)$ and initial reactive power $\bm{q}(0)$\\
\textit{Initialisation} : Initial control parameters $\bm{\varphi}_i =\{\bm{A}_i,k_i\}$ for all buses $i$\\
\textit{Computing graph} : Embedding the transition dynamics~\eqref{eq:closed-loop} in the computing graph through recurrent neural network
  \FOR {$epoch = 1$ to $I$}
  \STATE  Generate initial states $\bm{v}_h(0),\bm{q}_h(0)$ for the $h$-th batch, $h=1,\cdots,H$\\
  \STATE Run the controller in K steps and obtain the trajectory $\{\bm{v}_h(1),\bm{q}_h(1), \cdots, \bm{v}_h(T),\bm{q}_h(T)\}$
  \STATE  Calculate total loss of all the batches $Loss=\quad \sum_{h=1}^{H}\sum_{t=1}^{T} C_q(\bm{q}_h(t))+C_v(\bm{v}_h(t)-\mathbbold{1}_n)$\\
  \STATE  Update control paramters by passing $Loss$ to Adam optimizer:
  $\bm{\varphi} \leftarrow \bm{\varphi}-\alpha \text{Adam}(Loss)$
  \ENDFOR
 \end{algorithmic} 
 \end{algorithm}

\section{Case Study} \label{sec:simulation}

We verify the performance of the proposed approach on
IEEE 33-bus test feeders~\cite{baran1989network}.  We use TensorFlow 2.0 framework in Google Colab with a single Nvidia Tesla T4 GPU with 16GB memory to train the controllers.

The cost function that each controller collectively optimizes is $C(\bm{u})=\sum_{t=1}^{T}\left( ||\bm{v}_{t}||_1+\gamma||\bm{u}_t||_1\right)$, where $\gamma$ acts as a trade-off parameter and is set to be 0.001.  
The base unit for power and voltage is 100kVA and 12.66kV, respectively.   
The  bound  on  action $\Bar{\bm{u}}$  is generated to  be  uniformly  distributed  in $\text{uniform}[0.01, 0.05]\, \text{p.u.}$.
For training and testing the controller, we generate  500 trajectories by randomly setting the initial active and reactive power to be uniformly distributed in $\text{uniform}[0.3,1.7]\,\text{p.u.}$.

We compare the performance of the proposed adaptive controller~\eqref{eq:control_adpt} and standard linear control with $\hat{u}_i(v) = k_iv_i$ for all $i\in[n]$. Both the controllers are trained using the same algorithm in~\cite{cui2020}.  

\subsection{Illustrative example: Sinusoidal power injections}
For illustrative purposes, we demonstrate the performance of the proposed adaptive approach under sinusoidal time-varying oscillations of the net load. 
 The net load evolves as~\eqref{eq: p_t} where the basis $\phi_i(t) = sin(\eta_i t)$ and $\eta_i\sim \text{uniform}[0.003\pi, 0.008\pi]$. The coefficients $\bm{c}_i\sim \text{uniform}[0.05,0.25]$ for each bus $i$. 

\begin{figure}[ht]
\centering
\subfloat[Adaptive
]{\includegraphics[width=3.4
in]{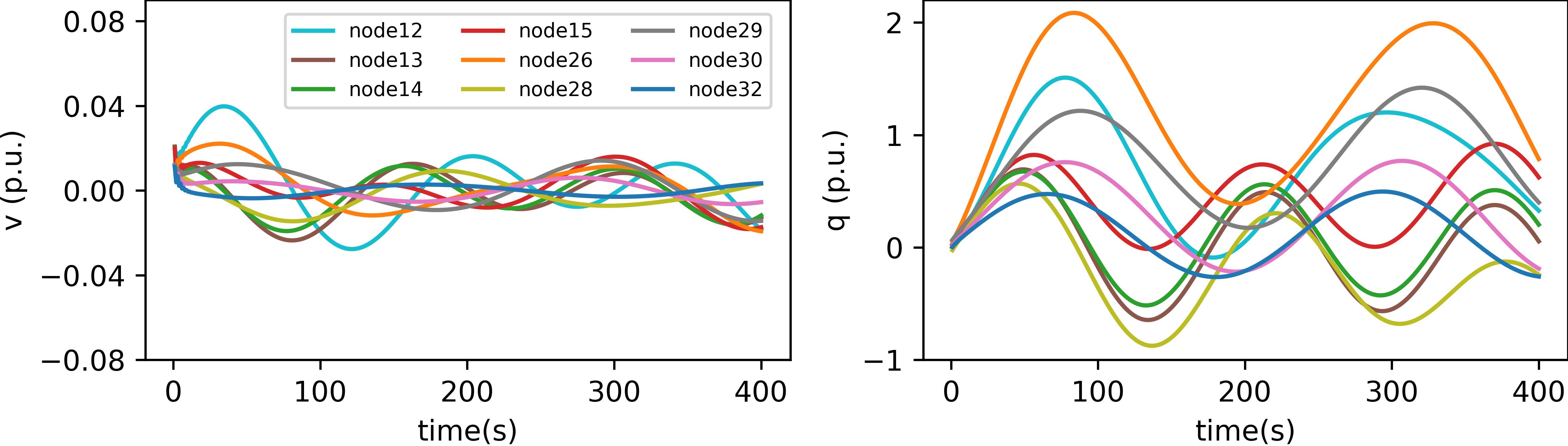}%
}
\hfil
\subfloat[Linear
]{\includegraphics[width=3.4in]{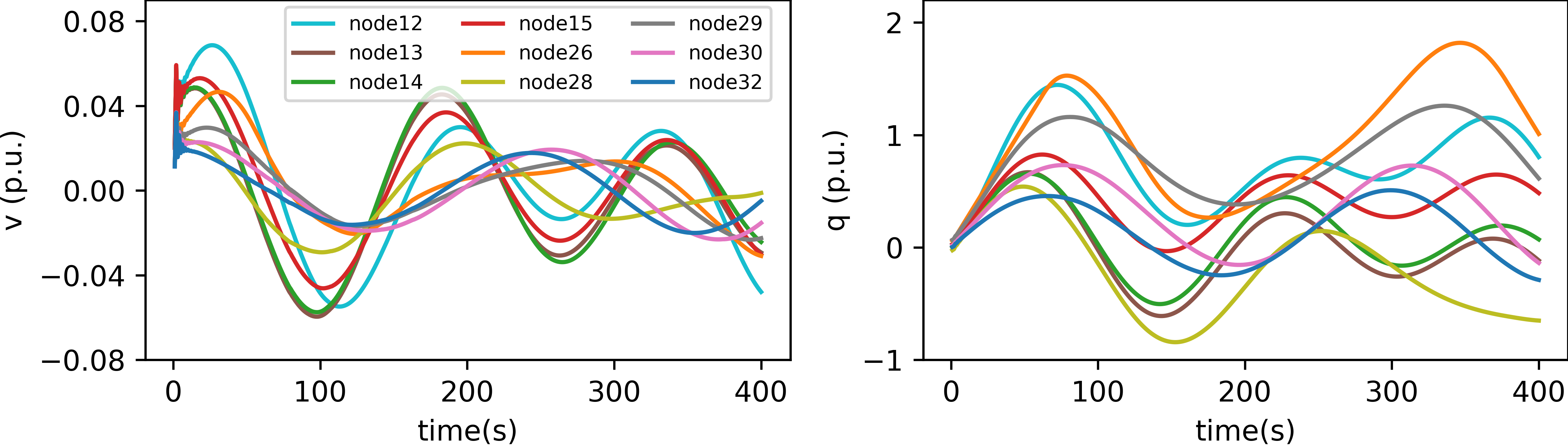}%
}
\caption{The voltage deviation $\Tilde{\bm{v}}$ and reactive power $\bm{q}$ with sinusoidal active power injections.  The adaptive approach achieves much lower voltage deviations and oscillations compared with the conventional incremental control law. }
\label{fig:Dynamic_sin}
\end{figure}

Figure~\ref{fig:Dynamic_sin} compares the voltage deviation and reactive power under the adaptive control and the linear control, respectively. The adaptive approach achieves significantly lower voltage deviation compared to the linear incremental control law. The performance improvement can be intuitively understood by examining the dynamics of the reactive power injection, $\bm{q}$. Because of the usage of predictions, the power injection under the adaptive approach more closely resembles a sinusoidal curve and thus compensates for the sinusoidal variations in the net load.

\subsection{Time-varying power injection from an operational distribution grid measurements}
\begin{figure}[ht]
\centering
\includegraphics[width=\columnwidth]{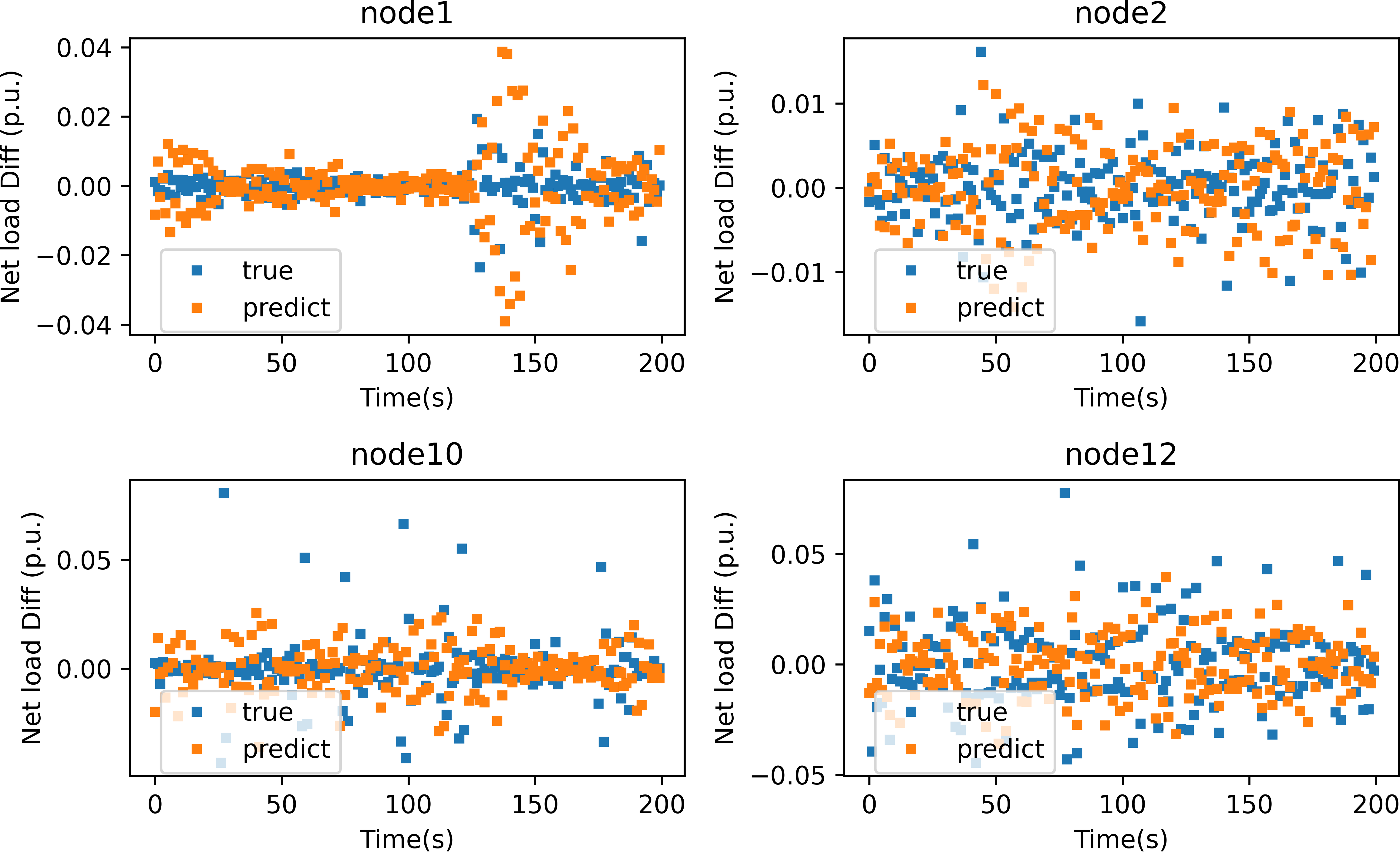}
\caption{ Time-varying active power injection
}
\label{fig:netload}
\end{figure}

To demonstrate the performance of the proposed method under real-world time-varying net load, we further conduct case studies using the measurements of active power output from an operational distribution grid in~\cite{xie2025digital}. 
As interpreted from equation~\eqref{eq: p_t}, the predictions target the difference in net load between neighboring time steps. Figure~\ref{fig:netload} shows the net load difference, where the blue dots represent the values calculated from true measurements and the orange dot indicates the prediction. While some prediction errors are present, the predictions can capture general trends in the time-varying load.

\begin{figure}[ht]
\centering
\includegraphics[width=0.7\columnwidth]{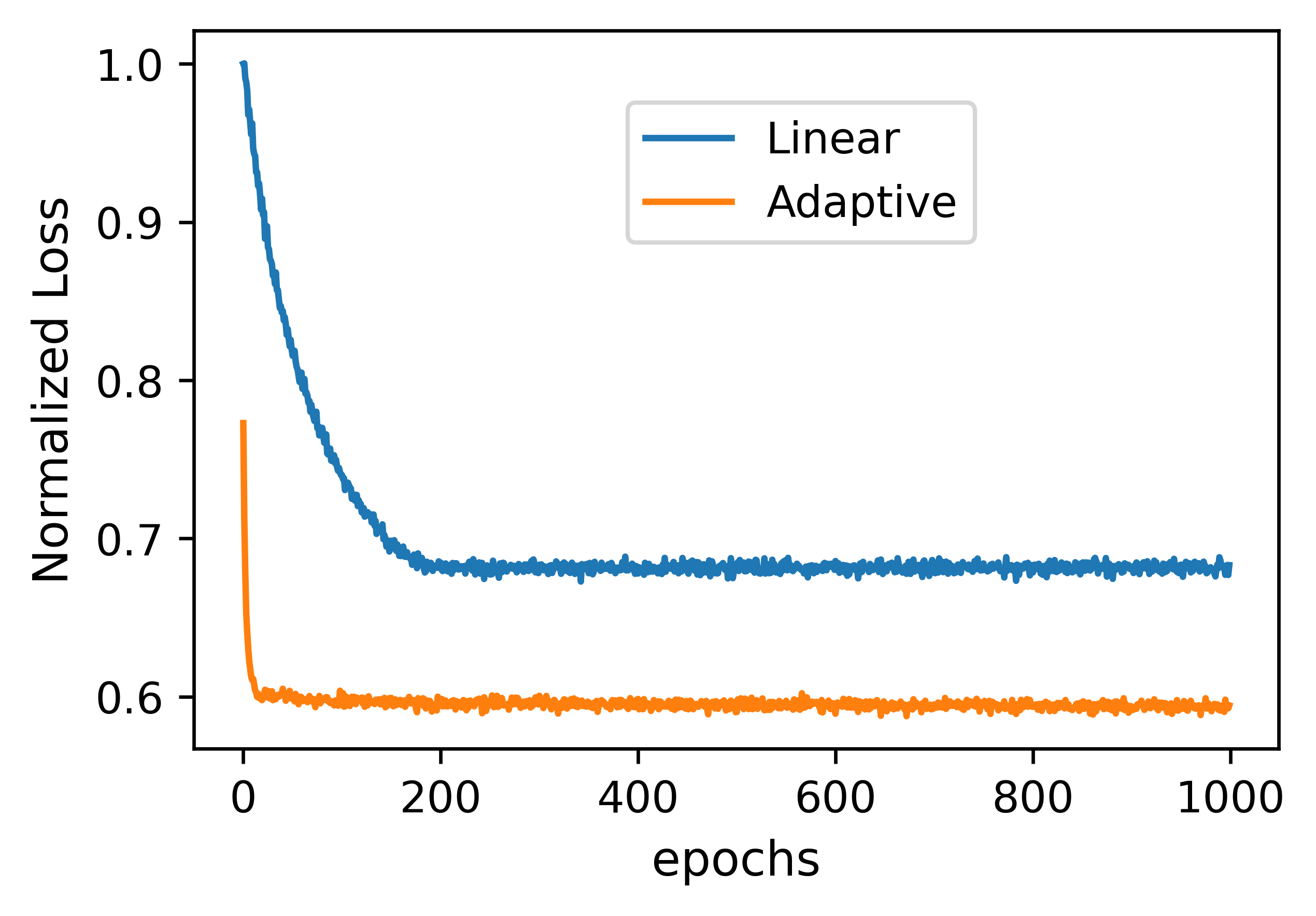}
\caption{ Average batch loss along epochs for IEEE-33 bus test case.  All converge,  with the adaptive approach achieving much lower loss than Linear control. 
}
\label{fig:loss_n10}
\end{figure}

The average batch loss during epochs of training is shown in Fig.~\ref{fig:loss_n10}. All the methods converge, with the adaptive approach having the lower loss after convergence.

\begin{figure}[ht]
\centering
\subfloat[Linear
]{\includegraphics[width=3.4in]{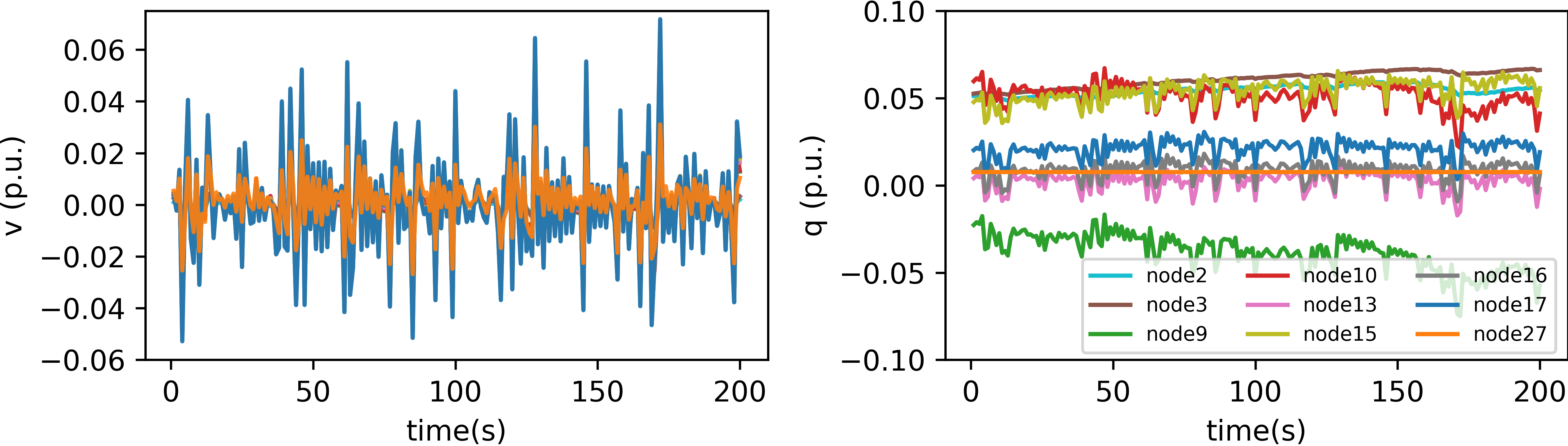}%
}
\hfil
\subfloat[Adaptive approach with predictions
]{\includegraphics[width=3.4
in]{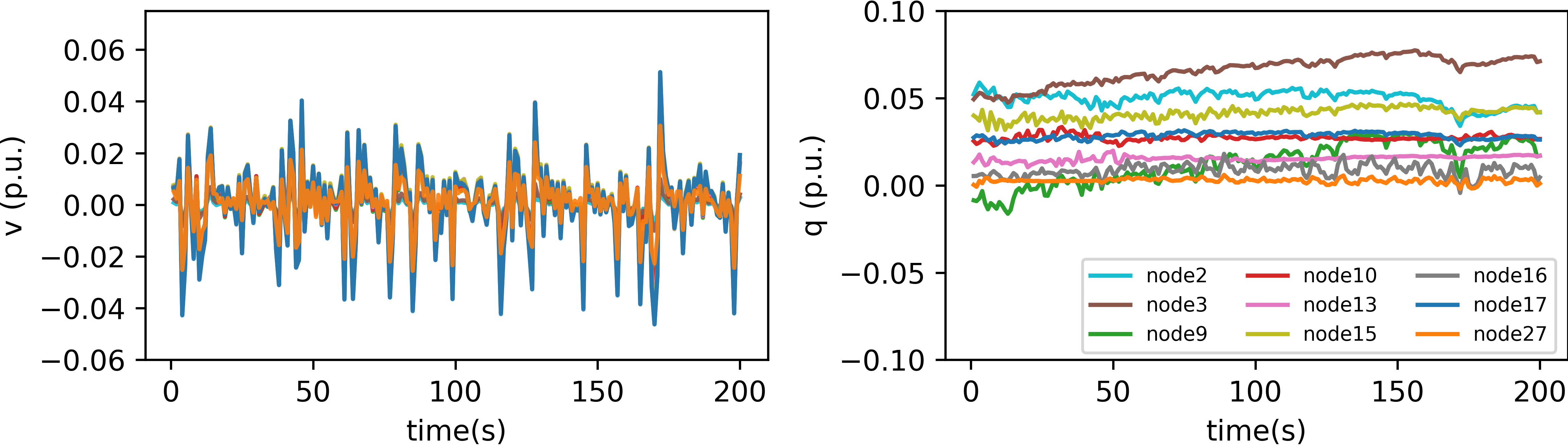}%
}
\hfil
\subfloat[Adaptive approach with exact local load information
]{\includegraphics[width=3.4
in]{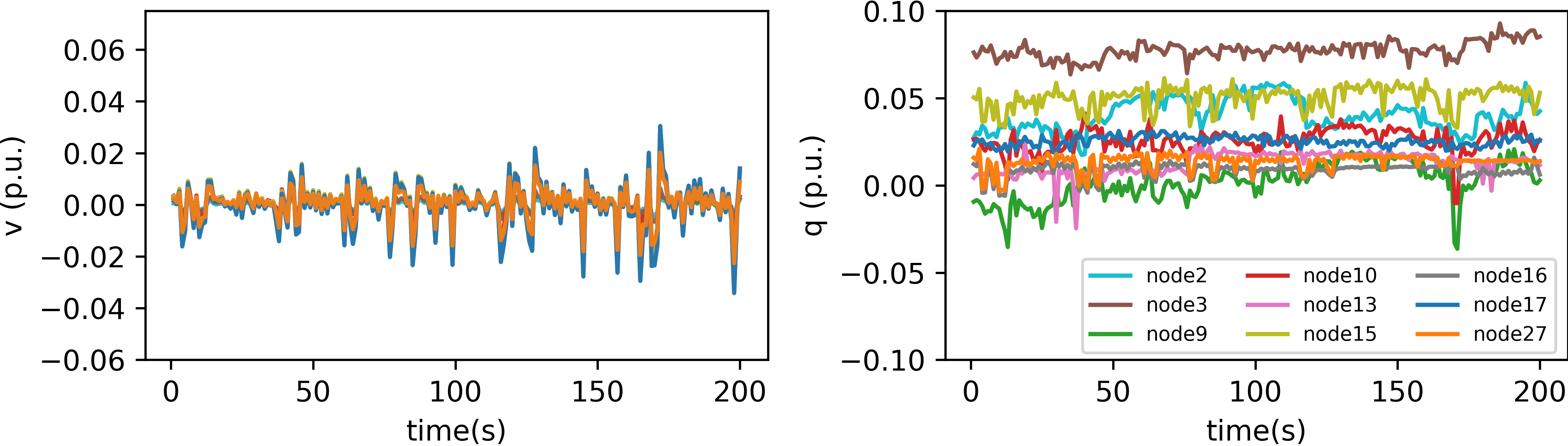}%
}

\caption{The voltage deviation $\Tilde{\bm{v}}$ and reactive power $\bm{q}$ with time-varying active power injections.  The adaptive approach achieves much lower voltage deviations and oscillations compared with conventional linear incremental control law. }
\label{fig:Dynamic_n10}
\end{figure}

\begin{figure}[ht]
\centering
\includegraphics[width=0.8\columnwidth]{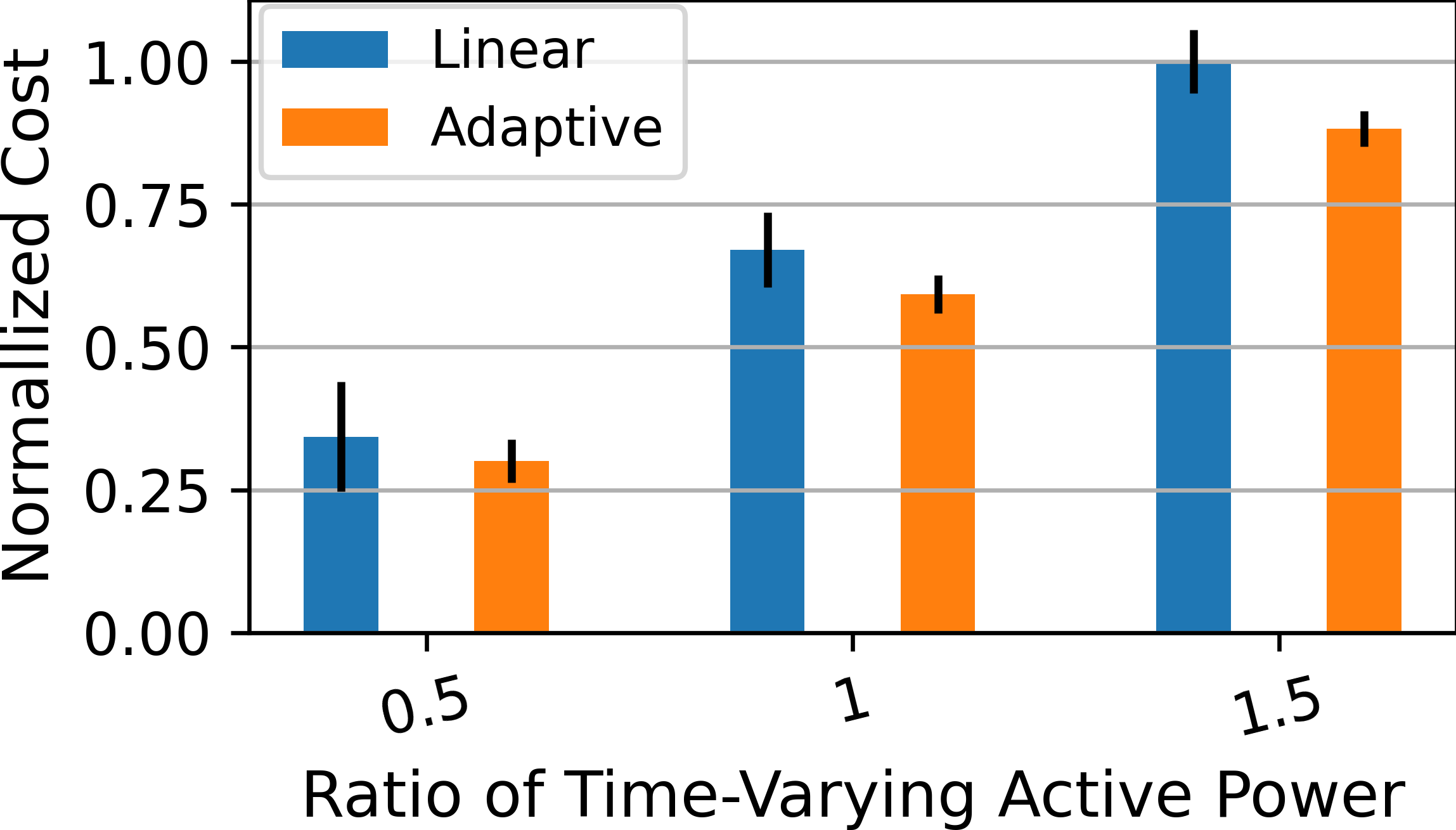}
\caption{ The average cost with error bar
on the randomly generated test set with size 100 and with different magnitudes of active power injection. The x-axis represents the ratio of the power injection magnitude in the test set relative to that in the training set.  The adaptive approach achieves a much lower cost compared with linear controllers for all the scenarios.
}
\label{fig:cost_n10}
\end{figure}

To demonstrate the performance of different controllers, Fig.~\ref{fig:Dynamic_n10} shows the dynamics of the voltage deviation $\bm{v}$ and the reactive power $\bm{q}$ under the same time-varying net load. The proposed adaptive approach using predictions in Fig.~\ref{fig:Dynamic_n10}(a) has lower voltage deviations compared with the linear controller in Fig.~\ref{fig:Dynamic_n10}(b), with a similar magnitude of reactive power. Therefore, the adaptive control law can greatly improve transient performances without incurring to high control cost. In addition, Fig.~\ref{fig:Dynamic_n10}(c) illustrates the dynamics under exact local load information, demonstrating that improved predictive accuracy can further reduce voltage deviations. This is consistent with the state convergence established in Theorem~\ref{thm:iss}.

As established in Theorem~\ref{lem: stability bounds}, the adaptive approach guarantees input-to-state stability by design. Thus, the controller stabilizes the system under time-varying net loads that differ from the training set, thereby achieving generalization to previously unseen scenarios.  To demonstrate this, Fig.~\ref{fig:cost_n10} compares the cost in the test set for $T=200$s, with different magnitudes of active power injection.  The x-axis represents the ratio of the power injection magnitude in the test set relative to that in the training set. At ratio = 1, the adaptive approach yields a cost approximately 9.77\% lower than that of the linear controller, consistent with the training performance.For ratio = 0.5 and ratio = 1.5, the adaptive approach achieves cost reductions of 10.59\% and 9.70\%, respectively. These results indicate that even for power injection levels not present in the training set, the proposed method consistently outperforms the linear controller. Therefore, the adaptive approach provides an efficient framework to improve control performances by leveraging predictions, which also achieves generalization across time-varying power injections not involved in the training set.

\section{Conclusion} \label{sec:conclusion} 
This paper proposes an adaptive voltage control framework for distribution systems with highly variable time-varying net load. We embed local load predictions as basis functions in voltage control and design adaptive combination coefficients to track evolving load conditions and prediction errors in real-time. This enables a significant reduction of voltage fluctuations with minimal modification to existing voltage control schemes. We proved that the closed-loop system achieves input-to-state stability with respect to prediction errors and model errors, and we explicitly derive the convergence of system states through time horizons. Case studies using both synthetic and real-world data from the campus grid demonstrate the effectiveness of the proposed approach in mitigating voltage fluctuations and violations.


\bibliographystyle{IEEEtran}
\bibliography{Reference}
\end{document}